# Dynamic MEMS-based optical metasurfaces

Chao Meng,[1†] Paul C. V. Thrane,[2†] Fei Ding,[1] Jo Gjessing,[2] Martin Thomaschewski,[1] Cuo Wu,[1,3], Christopher Dirdal,[2*] Sergey I. Bozhevolnyi[1*]

[1]Centre for Nano Optics, University of Southern Denmark, Campusvej 55, Odense DK-5230, Denmark.
[2]SINTEF Microsystems and Nanotechnology, Gaustadalleen 23C, 0737 Oslo, Norway.
[3]Institute of Fundamental and Frontier Sciences, University of Electronic Science and Technology of China, Chengdu 610054, China.

[†] These authors contributed equally to this work.
[*] Corresponding author emails: christopher.dirdal@sintef.no (C.D.); seib@mci.sdu.dk (S.I.B.)

**Abstract**
Optical metasurfaces (OMSs) have shown unprecedented capabilities for versatile wavefront manipulations at the subwavelength scale, thus opening fascinating perspectives for next generation ultracompact optical devices and systems. However, to date, most well-established OMSs are static, featuring well-defined optical responses determined by OMS configurations set during their fabrication. Dynamic OMS configurations investigated so far by using controlled constituent materials or geometrical parameters often exhibit specific limitations and reduced reconfigurability. Here, by combining a thin-film piezoelectric micro-electro-mechanical system (MEMS) with a gap-surface plasmon based OMS, we develop an electrically driven dynamic MEMS-OMS platform that offers controllable phase and amplitude modulation of the reflected light by finely actuating the MEMS mirror. Using this platform, we demonstrate MEMS-OMS components for polarization-independent beam steering and two-dimensional focusing with high modulation efficiencies (~ 50%), broadband operation (~ 20% near the operating wavelength of 800 nm) and fast responses (< 0.4 ms). The developed MEMS-OMS platform offers flexible solutions for realizing complex dynamic 2D wavefront manipulations that could be used in reconfigurable and adaptive optical networks and systems.

**INTRODUCTION**
Optical metasurfaces (OMSs) represent subwavelength-dense planar arrays of nanostructured elements (often called meta-atoms) designed to control local phases and amplitudes of scattered optical fields, thus being able to manipulate radiation wavefronts at a subwavelength scale (*1–5*). Numerous applications have already been demonstrated in the past decade, including free-space wavefront shaping (*6–9*), versatile polarization transformations (*10–13*), optical vortex generation (*14–16*), optical holography (*17–20*), to name a few. However, to date, most reported OMSs are static, featuring well-defined optical responses determined by OMS configurations that are set during fabrication. For more intelligent and adaptive systems, such as light detection and ranging (LIDAR), free-space optical tracking/communications and dynamic display/holography (*21–23*), it would be highly desirable to develop dynamic OMSs with externally controlled reconfigurable functionalities.

Realization of dynamic OMSs is very challenging because of the high density of array elements that are also arranged in nanometer-thin planar configurations. One of the currently investigated approaches relies on using dynamically controlled constituents, whose optical properties can be adjusted by external stimuli, thereby tuning their optical responses and reconfiguring the OMS functionalities. A variety of dynamic OMSs have been demonstrated by employing such materials, including liquid crystals (LCs) (*24–26*), phase-change materials (*27–31*), two-dimensional (2D) materials (*32–37*) and others (*38–41*). For example, by integrating the

OMS into a LC cell, reconfigurable beam steering was realized through electrically rotating the LCs in an addressable manner (*25*). Phase-change materials such as Ge$_2$Sb$_2$Te$_5$ (GST) (*27–30*) or VO$_2$ (*31*) were also employed to construct dynamic OMSs due to their reversible amorphous-crystalline or metal-insulator transitions. Furthermore, 2D materials, especially graphene, can be also used to implement dynamic OMSs since their optical properties can be remarkably adjusted through electrical gating/chemical doping with ultrafast switching speed, thus enabling dynamic OMSs with potentially ultrafast response (*32, 34*). Despite certain progress achieved with these configurations, there are still unresolved critical issues. Thus, LCs inherently require the polarization-resolved operation (*24–26*), phase-change materials feature relatively slow response times (*29–31*), while OMSs based on 2D materials suffer from relatively low modulation efficiencies (*35, 36*).

Another approach for realizing dynamic OMSs relies on direct modifying their geometrical parameters via mechanical actuations (*42–50*). Initial attempts include OMSs fabricated on elastomeric substrates with dynamic functionalities enabled by OMS stretching (*42, 45*). Faster and more accurate actuation can be achieved with micro-electro-mechanical systems (MEMS) that allow for electrically controlled actuation with nanometer precision and resolution, featuring also mature design and fabrication techniques (*43, 44, 46–50*). For example, varifocal lenses were realized with MEMS-actuated metasurface doublets, whose relative positions were controlled by MEMS actuators, resulting in continuous focal length tuning (*47*). In this configuration however, the two OMSs and their individual responses are not modified, making it difficult to use for dynamic wavefront manipulation in general. Very recently, through directly structuring OMSs on a movable silicon membrane of a silicon-on-insulator (SOI) wafer, dynamic one-dimensional (1D) wavefront shaping with fast response speed (~1 MHz) was demonstrated (*48*). In this case, direct OMS integration into the MEMS-actuated membrane leads to certain design limitations, resulting in polarization dependent performance and impeding implementation of 2D wavefront shaping.

Here, by combining a thin-film piezoelectric MEMS (*51–54*) with the gap-surface plasmon (GSP) based OMS (*6–8, 55*), we develop an electrically driven dynamic MEMS-OMS platform for realizing efficient, broadband and fast 2D wavefront shaping in reflection. The main idea is to split the conventional GSP based OMS (*6–8, 55*), so that an OMS layer containing metal nanobricks and back reflector are physically separated by an electrically controlled air gap, with an ultra-flat MEMS mirror serving as a moveable back reflector (Fig. 1A). Importantly, OMSs and MEMS mirrors are designed and fabricated in separate processing paths and then combined, ensuring thereby the design freedom on both sides and reducing the fabrication complexity. With this platform, we experimentally demonstrate dynamic polarization-independent beam steering and reflective 2D focusing (Fig. 1B). By electrically actuating the MEMS mirror and thus modulating the MEMS-OMS distance, polarization-independent dynamic responses with large modulation efficiencies are demonstrated. Specifically, when operating at a wavelength of 800 nm, the beam steering efficiency (in the +1$^{st}$ diffraction order) reaches 40% and 46% for the respective TM and TE polarizations (electric field parallel/perpendicular to the reflection plane, respectively), where 76% and 78% are expected from simulations. While the beam focusing efficiency reaches 56% and 53% (64% and 66% expected from simulations). Furthermore, the dynamic response of the investigated MEMS-OMSs is characterized with the respective rise/fall times of ~ 0.4/0.3 ms, characteristics that can be further improved by using MEMS mirrors optimized for bandwidth in the MHz range. For example, by using MEMS actuated membranes to ensure ~ 30 MHz switching speeds (*52–54*).

## RESULTS
### Operational principle
Similar to the conventional GSP based OMSs (*6–8, 55*), the proposed MEMS-OMS configuration represents a metal/insulator/metal (MIM) structure composed of a bottom thick gold layer atop a silicon substrate (MEMS mirror), an air spacer and a top layer with 2D arrays of gold nanobricks on a glass substrate (OMS structure). The air spacer gap $t_a$ can be finely adjusted by actuating the MEMS mirror (Fig. 2A). When the air gap is small ($t_a$ < 200 nm), the optical responses of OMS unit cells are determined by the GSP excitation and resonance in the MIM configuration (*55, 56*) and thus by nanobrick dimensions (*8, 55*). In order to progress further towards the design of dynamically controlled MEMS-OMSs, several geometrical OMS parameters must be determined. First, we set the operating wavelength at 800 nm and choose the OMS unit cell size of 250 nm that should be substantially smaller than the operating wavelength (*8, 55*). Assuming the smallest achievable air gap is between 20 – 50 nm, the nanobrick thickness $t_m$ is then optimized to achieve a wide phase coverage with large reflection amplitudes, resulting in the choice of $t_m$ = 50 nm (fig. S1). The nanobrick lateral dimensions, side lengths, are chosen to be equal to ensure the polarization-independent optical response. Analysis of the complex reflection coefficients of the OMS conducted for increased air gaps reveals that the phase gradient for different nanobrick side lengths progressively decreases, with the reflection phase and amplitude becoming independent on the nanobrick length at an air gap of $t_a$ = 350 nm (Fig. 2, B and C). This drastic transformation in the optical response is related to strong dependencies of the GSP excitation (at normal incidence) and GSP reflection at nanobrick terminations on the air gap: both decrease rapidly for increased air gaps, thereby attenuating and eventually eliminating the GSP resonance. The observed transformation of the reflection phase response (Fig. 2C) implies a simple and straightforward approach to realize dynamically controlled MEMS-OMSs: for a given smallest air gap (for example, 20 nm), one can design any conceivable GSP-based OMS (*57*), whose functionality can then be switched on and off by moving the MEMS mirror. Hereafter we demonstrate this approach by realizing dynamically controlled polarization-independent beam steering and reflective 2D focusing.

### Polarization-independent dynamic beam steering: design
The MEMS-OMS design for realizing dynamically controlled polarization-independent beam steering requires the choice of the number $N$ of unit cells in the OMS supercell that in turn determines the steering angle $\theta$ (in air) for the given unit cell size $\Lambda$ = 250 nm and light wavelength $\lambda$ = 800 nm: $\sin\theta = \lambda/N\Lambda$ (*6, 8, 55*). Bearing in mind experimental conditions, we chose an OMS supercell consisting of 12 cells so that the steering angle is $\theta$ = 15.5° in air, facilitating the characterization of well-separated $0^{th}/\pm1^{st}$ diffraction orders with a 20×/0.42 objective. Following the approach described above, the phase response calculated with the air gap $t_a$ = 20 nm for different nanobrick lengths is used to select the 12 nanobricks (Fig. 2C, red circles) and arrange them into an array along the $x$ direction (Fig. 2, D and E) to mimic the reflection coefficient of an ideal blazed grating: $r(x) = A\exp(i2\pi x/\Lambda_{sc})$ (*6, 8, 55*), where $A \leq 1$ is the reflection amplitude, and $\Lambda_{sc} = 12\Lambda$ is the grating (supercell) period. The available phase range at $t_a$ = 20 nm is slightly over 270° (the red dashed line in Fig. 2C), implying that it is impossible to design a supercell with 12 different unit cells ensuring a constant phase gradient (the latter requires the phase range of 11×30° = 330°). One possible approach to deal with this problem is to increase the phase (discretization) steps to 90°, so that the required phase range would decrease to 3×90° = 270°, resulting in the possibility to compose the supercell from duplicated ($\Lambda_{sc} = 4\Lambda \times 2 = 8\Lambda$) or triplicated ($\Lambda_{sc} = 4\Lambda \times 3 = 12\Lambda$) cells (*8*). Our simulations suggested another approach, in which two unit cells were left out empty, i.e., without nanobricks, while other 10 nanobricks cover the available phase range of 270°, thus ensuring better sampling of the phase profile and improving the efficiency of diffraction to the desired +1$^{st}$ order (fig. S2, A-F).

The reflected electric field (*x/y*-components) calculated for thus designed MEMS-OMS under TM/TE incident light at 800 nm wavelength with $t_a$ = 20 nm manifest smooth wavefronts travelling in the direction of the +1$^{st}$ diffraction order (Fig. 2F and fig. S2G). For increased air gaps, the phase gradients produced by the supercell nanobricks progressively decrease as expected (Fig. 2, B and C), with the phase gradient becoming zero and the reflected field returning to the specular reflection at an air gap of $t_a$ = 350 nm (Fig. 2G and fig. S2H). The associated decrease in the +1$^{st}$ order diffraction efficiency and increase in the 0$^{th}$ order one as a function of the air gap are practically linear, promising large modulation efficiencies available with the actuated MEMS-OMS (Fig. 2H). Thus, +1$^{st}$/0$^{th}$ order diffraction efficiencies are expected to change from 78%/0% to 0%/96% (for both TM and TE polarizations) when changing the air gap from 20 to 350 nm. The designed MEMS-OMS is expected to exhibit the broadband operation similar to that known for conventional GSP-based OMSs (*7*, *8*, *55*). Since the MEMS-OMS performance at large air gaps is equivalent to that of a mirror, its overall performance is then determined by that at the smaller air gap of $t_a$ = 20 nm, suggesting a 1-dB bandwidth of ~ 150 nm near the operating wavelength of 800 nm (Fig. 2I). As a final comment, it should be mentioned that, given the possibility of small air gap adjustments around the designed air gap of $t_a$ = 20 nm, the diffraction efficiencies for different wavelengths could be enhanced, thus improving the effective bandwidth of the MEMS-OMS device (fig. S2, I and J).

**Polarization-independent dynamic beam steering: characterization**
The MEMS-OMS for polarization-independent beam steering designed above (Fig. 2) was assembled from a separately fabricated OMS, an ultra-flat MEMS mirror (*51*) and a printed circuit board (Fig. 3A, for details see Materials and Methods along with Supplementary fig. S3, A-C). The possibility of fabricating the MEMS mirror and OMS separately simplifies the design and fabrication processes, for example by allowing the two components to be produced in separate processing lines with different minimum line width capabilities. The fabricated MEMS mirror and OMS were characterized individually using optical and scanning electron microscopes (Fig. 3, B and C). When joining the MEMS mirror and OMS it is important to avoid any particles that can obstruct the mirror from getting close enough to the OMS. Because the mirror [i.e., 3 mm in diameter] was much larger than the OMS [i.e., 30 × 30 μm$^2$ in size], the OMS was fabricated on top of a 10-μm-high pedestal, the idea being that any particles smaller than 10 μm outside the pedestal will not prevent the OMS and MEMS mirror from coming into contact. This pedestal did not affect the fabrication of the nanobricks, featuring overall consistency with the design apart from slightly rounded corners and minor size deviations (Fig. 3C). After assembling the MEMS-OMS, the MEMS-OMS separation was estimated using white light interferometry (Zygo NewView 6000) to be ~ 2 μm (fig. S3D), which is well within the ~ 6-μm-large moving range of the MEMS mirror (see Materials and Methods along with Supplementary fig. S3E). Following that, we estimated the smallest achievable separation between the MEMS mirror and OMS substrate surface (crucial for efficient modulation) by using a multi-wavelength interferometry (fig. S4). We found by actuating the MEMS mirror that, for several assemblies, this gap ($t_m$ + $t_a$) can be as small as ~100 nm (fig. S4), corresponding to $t_a$ ~ 50 nm, and these samples were then selected for further optical characterizations.

To characterize the MEMS-OMS performance, we used a wavelength-tunable (~ 700 – 1000 nm) laser with the corresponding optical, polarization and imaging components (see Materials and Methods along with Supplementary fig. S5). The MEMS mirror is electrically actuated to modulate the optical response of the MEMS-OMS observed visually in both direct object (OMS surface) and Fourier image planes (Fig. 4A). In the direct object images, this effect of power redistribution is seen in the appearance (at non-zero actuation voltages) of well-pronounced interference fringes formed due to the interference between the residual specular reflection and the +1$^{st}$ order diffracted beam. For both polarizations, the redistribution of radiation power between the 0$^{th}$ and +1$^{st}$ diffraction orders are well pronounced, reaching the maximum contrast at 3.75 V with the diffraction efficiencies of 40%/46% for the respective TM/TE polarizations (Fig. 4, A and B). The high-contrast dynamic beam steering, induced by actuating the MEMS mirror with the alternating voltages of 0 and 3.75 V at a slow switching speed, is clearly seen in the movie captured by the CCD camera (Supplementary Movie S1). The MEMS-OMS operation is found polarization-independent and broadband, exhibiting the 1dB bandwidth of ~ 150 nm (Fig. 4C). By actuating the MEMS mirror with a periodic rectangle signal and detecting the spatially separated 0$^{th}$/+1$^{st}$ order diffraction fields, one observes relatively fast switching with the rise/fall times of ~ 0.4/0.3 ms, respectively (Fig. 4D). The response speed is related to the intrinsic oscillation frequency of the MEMS mirror, thus being dependent on the MEMS design parameters such as geometry, weight, stiffness and so on (*51–54*). Note that the standard thin-film MEMS mirror used is rather large (~ 3 mm in diameter, Fig. 3A) with its surface area orders of magnitude larger than that of the OMS area (~ 30 × 30 $\mu m^2$ in size, Fig. 3B), considerably slowing down the dynamic response. Bearing in mind the possibility of optimizing the MEMS mirror for fast switching speeds, one should expect reaching operation bandwidths in the MHz range, indeed, current state of the art in thin-film piezoelectric MEMS can achieve ~ 30 MHz switching frequencies (*52–54*).

Concluding the presentation of the demonstrated MEMS-OMS for polarization-independent beam steering, we would like to note that, although the experimentally observed performance (Fig. 4) is somewhat inferior to that expected from our simulations (Fig. 2), the experimental performance can be improved. The deterioration can be attributed partly to fabrication imperfections and to the smallest air gap $t_a$ that was achieved in practice. It seems that the air gap decreases with applying the actuation voltage only up to ~ 3.75 V, resulting thereby in increasing +1$^{st}$ and decreasing 0$^{th}$ order diffraction efficiencies, whereas for larger voltages the MEMS mirror starts to move slightly away from the OMS, probably because of the residual contaminants on the substrate or bending at the pedestal edges which prevent the MEMS mirror from moving further closer to the OMS surface. Both better fabrication accuracy and smaller air gaps are feasible and expected to be realized in further experiments.

**Polarization-independent dynamic 2D focusing: design**
The MEMS-OMS design for realizing dynamically controlled polarization-independent 2D beam focusing in reflection requires the choice of diameter *D* and focal length *f* of the OMS lens that in turn determines the numerical aperture *NA* for the given refractive index in the image space $n$ = 1.46 at an incident wavelength of $\lambda$ = 800 nm: $NA = n\sin[\tan^{-1}(D/2f)]$. To realize strong focusing, we chose *D* = 14 $\mu$m and *f* = 15 $\mu$m, so that $NA \approx 0.62$ is expected, which should be adequate to enable high-efficiency reflective 2D focusing (*7*). Following the same design approach used in demonstrating MEMS-OMS for dynamic beam steering, we use the phase response calculated with air gap $t_a$ = 20 nm for different nanobrick lengths (the red dashed line in Fig. 2C) to extract the proper unit cells and arrange them into a circular region with *D* = 14 $\mu$m (Fig. 5A), approximating a hyperboloidal phase profile (*7, 9*) : $\phi_{2D} = \frac{2\pi}{\lambda}n(f - \sqrt{x^2 + y^2 + f^2})$ in the *xy*-plane (Fig. 5B). The above phase profile is also discretized with the step size $\Lambda$ = 250 nm along both *x/y*- directions, matching the unit cell size ($\Lambda$ = 250 nm). In contrast to the previous work, we do not limit the choice of unit cells to a discrete design space [i.e., unit cells with discrete phase steps of 45° (*7*)]. Instead, appropriate lengths of the nanobricks are chosen from the entire space of simulation results (the red dashed line in Fig. 2C), thus ensuring better sampling of the 2D phase profile with minor deviations (fig. S6, A and B) from the required one (Fig. 5B). The deviation between the required and available phase profiles results mostly from the achievable phase coverage of ~ 270°, a limitation that could be circumvented by including more complex unit cell elements such as detuned GSP resonators (*58*).

Bearing in mind high computational demands when simulating 2D focusing (and thus aperiodic) OMSs, we estimate the focusing performance by simulating the corresponding (reduced to a 1D aperiodic configuration) OMS (fig. S6, C and D), which is designed to provide a 1D hyperboloidal phase profile $\phi_{1D} = \frac{2\pi}{\lambda} n(f - \sqrt{x^2 + f^2})$, while the $D$, $f$ and $\lambda$ are the same as that of the above-designed OMS with the 2D phase profile. The reflected intensity distributions calculated for this simplified MEMS-OMS under TM/TE incident light at 800 nm wavelength with $t_a$ = 20 nm manifest high focusing quality with a diffraction-limited spot situated at the focal length of ~ 15 $\mu$m (Fig. 5C and fig. S6E). For increased air gaps, the phase gradients produced by nanobricks with different lengths progressively decrease (Fig. 2, B and C), approaching zero at an air gap of 350 nm with the reflection transformed into specular reflection (Fig. 5D and fig. S6F). The associated reflected electric fields calculated near the focus display smoothly converging and planar wavefronts at air gaps $t_a$ = 20 (Fig. 5E and fig. S6G) and 350 nm (Fig. 5F and fig. S6H), respectively, implying high-efficiency operation of the actuated MEMS-OMS. Taking into account the possibility of adjusting the air gap to maximize the focusing efficiency at other (than the design) wavelengths, we evaluated the focusing efficiencies achievable at different wavelengths with varied air gaps (Fig. 5, G and H). The maximum achievable focusing efficiencies at the design wavelength of 800 nm are estimated to be ~64%/66% (TM/TE) for the air gap of ~ 20 nm as expected. For other wavelengths, the polarization-independent focusing behavior is well-maintained, while the corresponding maximal focusing efficiencies are expected to achieve at slightly different air gaps. To better visualize this feature, the focusing efficiency as a function of the air gap is explicitly plotted for distinct wavelengths of 750, 800 and 950 nm (Fig. 5I), showing for all wavelengths a nearly linear decrease of the efficiency for increasing air gaps without noticeably changes in the reflected field distributions (fig. S6, I-L).

**Polarization-independent dynamic 2D focusing: characterization**
The MEMS-OMS for polarization-independent dynamic reflective 2D focusing designed as described above (Fig. 5) was assembled following the fabrication and pre-characterization processes similar to those employed when assembling the dynamic beam steering MEMS-OMS. Optical microscopy and SEM are employed for monitoring the possible contaminants on the OMS surface as well as the fabrication quality (the upper-left inset of Fig. 6A and fig. S7).

To characterize the dynamic focusing MEMS-OMS, we electrically actuated the MEMS mirror and observed corresponding optical responses in the direct object plane (Fig. 6B). Since the MEMS-OMS was designed to exhibit a very short focal length of ~ 15 $\mu$m, it was not possible to directly access the focal plane using a beam splitter and a low-divergent incident laser beam. Instead, the focusing effect was verified by illuminating the MEMS-OMS with a focused incident beam and placing the MEMS-OMS surface plane B at a distance of ~ $2f$ (the double focal length of the MEMS-OMS) away from the incident beam focal plane A (see inset in Fig. 6A). According to the ray optics, the beam reflected by the OMS (when close to the MEMS mirror) will then be focused again at the focal plane of the objective (plane A in the bottom-right inset of Fig. 6A). At the same time, the reflection from the unstructured substrate surface (outside the OMS area) would be strong diverging (see the bottom-left inset in Fig. 6A). If one moves the MEMS-OMS surface to plane A, the reflection behavior will be reversed: the reflection by the OMS will be diverging (after the objective) and that by the unstructured surface – collimated. This procedure was successfully used and described in detail in the previous experiment conducted with the static focusing OMS (*7*). In the current case with the dynamic focusing MEMS-OMS, it is expected for the MEMS-OMS arrangement to switch between the focusing configuration, when the applied voltage would bring the OMS very close to the MEMS mirror, and the mirror reflecting configuration for relatively small applied voltages that would correspond to sufficient large OMS and MEMS mirror separations.

To observe this transformation, we monitored the reflected light from the MEMS-OMS positioned at plane B while actuating the MEMS mirror. For both polarizations, the switching of the reflected light between the mirror (at $V_{b1}$ = 10.00 V) and focusing (at $V_{b2}$ = 14.50 V) cases was clearly visualized (Fig. 6B), with the focusing efficiencies reaching their maxima of ~56%/53% at $V_{b2}$ = 14.50 V for the respective TM/TE light incidence at the wavelength of 800 nm (Fig. 6A). At the same time, the reflection from the unstructured substrate surface was not influenced with the applied voltages, revealing however that the reflection from the substrate at plane A is strikingly similar to the TM/TE reflection from the OMS at plane B with the applied voltage being $V_{b2}$ = 14.50 V (Fig. 6B). The latter evidences a rather high efficiency and excellent quality of polarization independent focusing by the MEMS-OMS at $V_{b2}$ = 14.50 V. The dynamic evolution of the reflected field from the MEMS-OMS positioned at plane B, induced by actuating the MEMS mirror with stepwise increased voltages from 10.00 to 14.50 V, is clearly observed with a CCD camera (Supplementary Movie S2). Due to the usage of the same MEMS component as that in the dynamic beam steering MEMS-OMS, similar response time of ~ 0.4 ms is expected. It is finally worth noting that, according to the current state of the art in thin-film piezoelectric MEMS techniques (*52–54*), MEMS-OMS components with ~ MHz switching bandwidth should be feasible and expected for further developments.

**Discussion**
We have developed the electrically driven dynamic MEMS-OMS platform by combining a thin-film piezoelectric MEMS mirror with a GSP-based OMS. This platform offers controllable phase and amplitude modulation of the reflected light by finely actuating the MEMS mirror. We have designed and experimentally demonstrated MEMS-OMS devices operating in the near-infrared wavelength range for dynamic polarization-independent beam steering and reflective 2D focusing, both exhibiting efficient (~ 50%), broadband (~ 20% near the operating wavelength of 800 nm) and fast (< 0.4 ms) operation. The operation of both devices relies on the phase response transformation when changing the MEMS-OMS separation. The same operation principle can be used to design a MEMS-OMS for dynamically controlling any functionality available for conventional GSP-based OMSs, from polarization control/detection to vector/vortex beam generation (*57*): for a given smallest air gap, one designs the GSP-based OMS exhibiting a required functionality that can then be switched on and off by moving the MEMS mirror towards and away from the OMS surface.

Moreover, the nontrivial modification of the size-dependent phase response with the MEMS-OMS separation (Fig. 2B), which can accurately be adjusted by electrical MEMS actuation, suggests a possibility of realizing more sophisticated dynamic functionalities. Thus, we have also designed and experimentally demonstrated the MEMS-OMS device for polarization-independent dynamic beam steering between three (0$^{th}$, 1$^{st}$ and 2$^{nd}$) diffraction orders, corresponding to reflection angles of 0º, 7.7º and 15.5º in air under normally incident light with 800 nm wavelength. The OMS configuration (fig. S8) consisted of two OMSs with different supercells with $\Lambda_{sc1}$ = 12$\Lambda$ and $\Lambda_{sc2}$ = 24$\Lambda$ optimized at two distinct air gaps and interleaved by adopting a random-interleaving strategy (*59*). The experimental characterization (fig. S9) has confirmed the intended dynamic MEMS-OMS response: with the actuation voltage increasing, the first and second diffraction orders became visible, succeeding one another, in accordance with our simulations (fig. S8I).

Another promising direction for further research and development is to circumvent the necessity of bringing the MEMS mirror very close (~100 nm) to the OMS surface. For large MEMS-OMS separations, one can make use of localized plasmon resonances due to excitation of short-range surface plasmon-polaritons (SR-SPPs) in thin metal films (*56*). Our preliminary simulations showed that the SR-SPPs resonances hybridize with the Fabry-Pérot resonances (supported with wavelength-large air gaps) (*60, 61*) and open a similar to the considered above route to modify the OMS phase response by controlling the air gap (fig. S10, A-C). With this approach the MEMS-OMS can be operated near the air gap of ~ 1 μm or more, as demonstrated with our simulations of dynamic beam steering (fig. S10, D-L), thus avoiding the problem of realizing nm-sized air gaps. Overall, we believe that diverse functionalities with dynamically reconfigurable performances can be realized using the developed MEMS-OMS platform, thus opening fascinating perspectives for successful realization of high-performance dynamically controlled devices with potential applications in future reconfigurable/adaptive optical systems.

**Materials and Methods**
**Simulation methods**
All numerical simulations were performed using COMSOL Multiphysics 5.5. We modeled one individual Glass/Au/Air/Au unit cell (Fig. 2A), where periodic boundary conditions were applied in both x/y- directions, and linearly x-polarized light at the design wavelength of 800 nm was normally incident onto the unit cell from the upper glass layer. The permittivity of Au is described by the interpolated experimental values (62), and the glass layer is taken as a lossless dielectric with a constant refractive index of 1.46. Then the complex reflection coefficients (Fig. 2B) were calculated as a function of nanobrick lengths $L_x$, and air gap $t_a$ with other parameters being as follows: $\lambda$ = 800 nm, $t_m$ = 50 nm, $\Lambda$ = 250 nm, and $L_y = L_x$ to ensure the polarization-independent optical responses.

To design the MEMS-OMS for dynamic beam steering, the phase response calculated with the air gap $t_a$ = 20 nm for different nanobrick lengths is used to select the lengths of 12 nanobricks (Fig. 2C) for approximating the reflection coefficient of an ideal blazed grating: $r(x) = A\exp(i2\pi x/\Lambda_{sc})$ (6, 8, 55), where $A \leq 1$ is the reflection amplitude, and $\Lambda_{sc} = 12\Lambda$ is the grating (supercell) period. Reflected light directed to different diffraction orders are monitored, with different air gaps $t_a$ and incident wavelengths $\lambda$ for estimating the dynamic diffraction efficiencies and operation optical bandwidths, respectively (Fig. 2, D-I and fig. S2, D-J).

MEMS-OMS for dynamic beam focusing is designed and simulated in a similar fashion. Nanobricks from the phase response calculated with the air gap $t_a$ = 20 nm for different nanobrick lengths (Fig. 2C) are selected to approximate a 1D hyperboloidal phase profile of $\phi_{1D} = \frac{2\pi}{\lambda} n(f - \sqrt{x^2 + f^2})$ (7, 9) within a 14-μm-diameter region in the xy-plane (fig. S6 C and D). Reflected fields are monitored to visualize the dynamic beam focusing and estimate corresponding focusing efficiencies as a function of the gap sizes $t_a$ and incident wavelengths $\lambda$, for both TM/TE polarizations (Fig. 5, C-I and fig. S6, E-L).

**Fabrication and assembly of the MEMS-OMS devices**
The OMSs for developing MEMS-OMS for dynamic beam steering/focusing were fabricated using standard electron-beam lithography (EBL), thin-film deposition and lift-off techniques. First, a 100 nm thick poly(methyl methacrylate) (PMMA, 2% in anisole, Micro Chem) layer and a 40 nm thick conductive polymer layer (AR-PC 5090, Allresist) were successively spin-coated on a 16 × 16 mm² glass substrate (Borofloat33 wafer, Wafer Universe). Note that the glass substrate was pre-processed to have a 10-μm-high circular/cross-shaped pedestal on one side using optical lithography and wet etching. The OMSs were then defined on the pedestal of the glass substrate using EBL (JEOL JSM-6500F field emission scanning electron microscope with a Raith Elphy Quantum lithography system) and subsequently developed in 1:3 solution of methyl isobutyl ketone (MIBK) and isopropyl alcohol (IPA). After development, a 1 nm Ti adhesion layer and a 50 nm Au layer were deposited using thermal evaporation. The Au nanobricks were finally formed atop the pedestal on the glass substrate after a lift-off process (Fig. 3 and fig. S7). Owing to the large size of the MEMS mirror (~ 3 mm in diameter) in comparison to the OMS (30×30 μm² in size), the pedestal on the glass substrate is very practical for reducing the possible contaminants between the MEMS mirror and OMS surface, thus promising high-efficiency modulation of the MEMS-OMS devices.

The MEMS mirror, which is very similar to the previously reported ultra-planar, long-stroke and low-voltage piezoelectric micromirror (*51*), is fabricated using standard semiconductor manufacturing processes (fig. S3A) and incorporating thin film lead zirconate titanate (PZT) for actuation. First, a platinum (Pt) bottom electrode, a 2-μm-thick PZT film and a top electrode consisting of TiW/Au were deposited on a silicon-on-insulator (SOI) wafer (fig. S3A, first panel). Then a central circular aperture of 3 mm was opened by using deep reactive ion etching (DRIE) of silicon and etching of the buried oxide (fig. S3A, second panel). An annulus trench is etched into the backside of the wafer, thereby releasing the circular plate. Finally, the wafer backside is sputtered with Au (fig. S3A, third panel) for acting as the ultra-flat MEMS mirror that is of vital importance in developing dynamic MEMS-OMS.

After the fabrication of both OMS and MEMS mirror, we move to the assembly and packaging processes for making MEMS-OMS devices (fig S3A, fourth panel). Before assembly, the surface topography of the MEMS mirror and glass substrate were measured by a white light interferometry (Zygo NewView 6000), so as to select favorable areas on both sides with the least amount of contaminants and surface roughness that might obstruct the MEMS mirror from getting close enough to the OMS. Then, the MEMS mirror was glued to the glass substrate upon which the OMS has been structured (fig. S3, B and C). The spacing between the MEMS mirror and the glass substrate ($t_a+t_m$) was measured to be commonly ~ 2 μm after mounting (fig. S3D), well within the 6-μm moving range of the MEMS mirrors (fig. S3E). Finally, the MEMS-OMS was glued to a printed circuitry board (PCB), and gold wire bonding is used to connect electrically to the MEMS electrodes for enabling simple connection to a voltage controller used to actuate the MEMS mirror.

After the MEMS-OMS assembly, we applied multi-wavelength interferometry to estimate the smallest achievable separation between the MEMS mirror and OMS surface (fig. S4). We found by actuating the MEMS mirror that, for several assemblies, this gap ($t_m+t_a$) can be as small as ~ 100 nm corresponding to $t_a$ ~ 50 nm, and these samples were then selected for further optical characterizations.

**Optical characterization of MEMS-OMS**

To characterize the performances of the MEMS-OMS for dynamic 2D wavefront shaping, we used a fiber-coupled wavelength-tunable Ti:sapphire laser (Spectra Physics 3900S, wavelength range: 700~1000 nm), whose light was directed through a half-wave plate (HWP, AHWP05M-980, Thorlabs), a Glan-Thomson polarizer and a first beam splitter ($BS_1$, BS014, Thorlabs) successively, and then focused by an objective (Obj, M Plan Apo, 20×/50× magnifications, Mitutoyo) onto the MEMS-OMSs. The reflected light was collected by the same objective and directed via $BS_1$ and a second BS ($BS_2$, BS014, Thorlabs) to two optical paths terminated with two CCD cameras (DCC1545M, Thorlabs) for visualizing respective direct object and Fourier plane images (fig. S5). Note that the objective of 20×/0.42 and 50×/0.55 are employed for measuring respective MEMS-OMS for dynamic beam steering and focusing.

During the measurement, the MEMS mirror was electrically actuated to modulate the optical responses of the MEMS-OMS devices. In measuring MEMS-OMS for dynamic beam steering, Fourier plane images are captured for the incident beam on the OMS area and the unstructured substrate (as a reference) for estimating diffraction efficiencies. The response time of the MEMS-OMS was evaluated by actuating the MEMS mirror with a periodic rectangle signal from a function generator (TOE 7402, TOELLNER). The spatially separated $0^{th}/1^{st}$ diffraction orders at the Fourier plane could be selected by an iris and then projected to a photodetector (PDA20CS-EC, Thorlabs), which was connected to an oscilloscope (DSOX2024A, Keysight) for visualizing and recording the corresponding modulated signals.

**Acknowledgments**
**Funding:** This project has received funding from the ATTRACT project funded by the EC under Grant Agreement 777222, the University of Southern Denmark (SDU2020 funding), the VKR Foundation (Award in Technical and Natural Sciences 2019 and Grant No. 00022988) and the EU Horizon 2020 research and innovation programme (the Marie Sklodowska-Curie grant agreement No. 713694).
**Author contributions:** C.D. and S.I.B. conceived the idea. C.M. and F.D. performed the numerical simulations. C.M. and C.W. fabricated the OMS samples, P.C.V.T. and J.G. fabricated the MEMS mirror and assembled MEMS-OMS devices, C.M. and P.C.V.T. conducted the optical measurements. C.M. and M.T. performed the response time measurement. All authors contributed to the manuscript writing. C.D. and S.I.B. supervised the project.
**Competing interests:** The authors declare that they have no competing interests.
**Data and materials availability:** All data needed to evaluate the conclusions in the paper are present in the paper and/or the Supplementary Materials. Additional data related to this paper may be requested from the authors.

# Figures and Tables

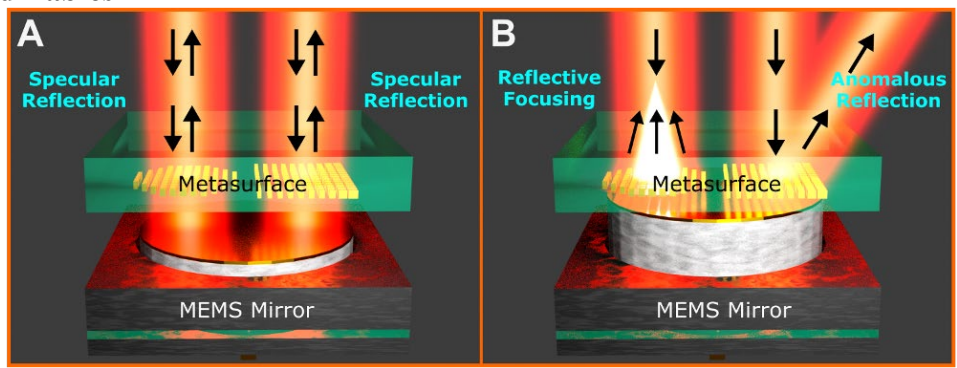

**Fig. 1. 2D wavefront shaping with the MEMS-OMS.** (**A**) Schematic of mirror-like light reflection by the MEMS-OMS before the actuation, i.e., with the initial gap of ~ 350 nm between the OMS nanobrick arrays and MEMS mirror. Incident light is specularly reflected by the MEMS-OMS regardless the OMS design. (**B**) Schematic of demonstrated functionalities, anomalous reflection and focusing (depending on the OMS design), activated by bringing the MEMS mirror close to the OMS surface, i.e., by decreasing the air gap to ~ 20 nm.

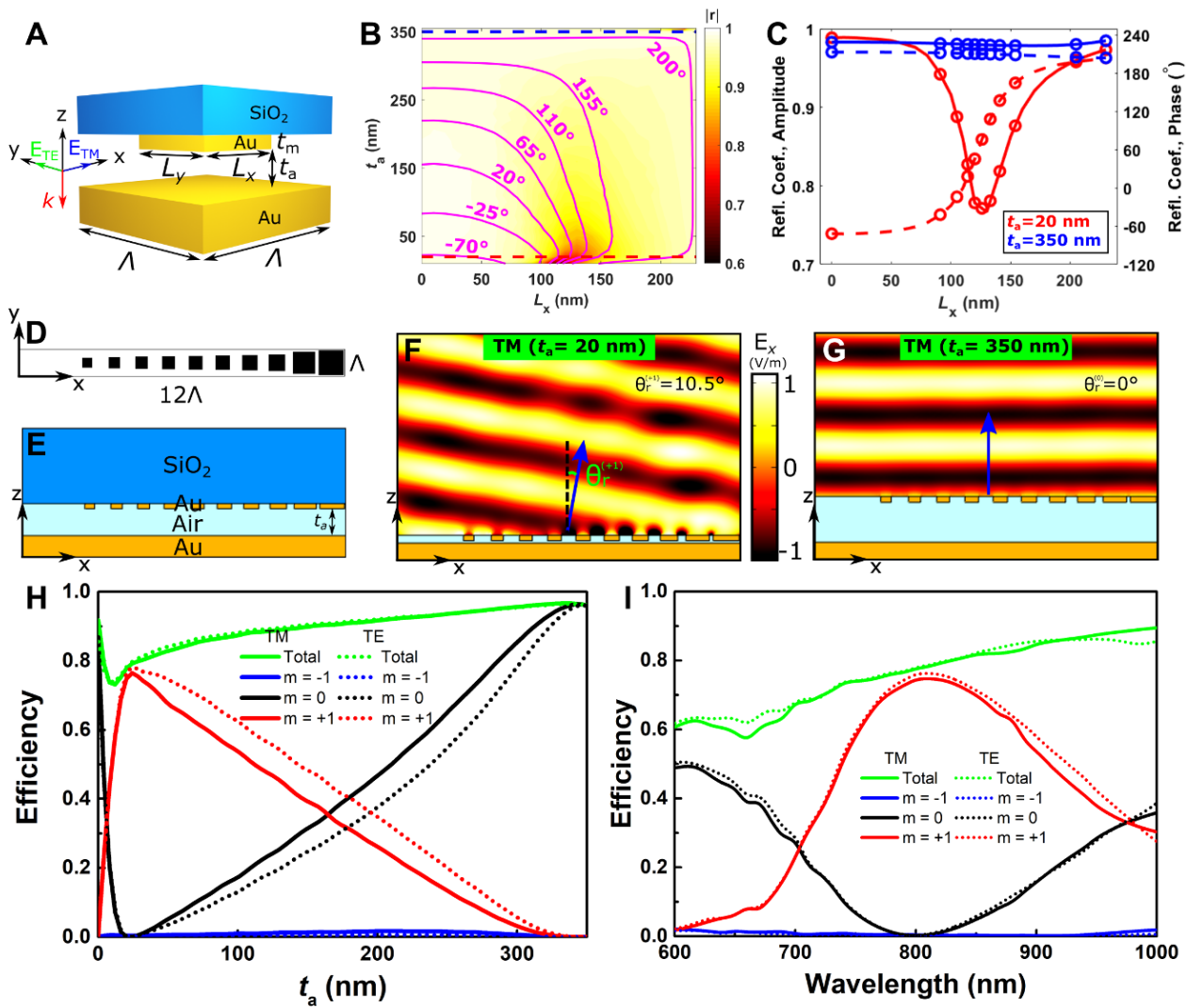

**Fig. 2. Polarization-independent dynamic beam steering: design.** (**A**) Schematic of the OMS unit cell including the air gap and gold mirror. (**B**) The complex reflection coefficient $r$ calculated

as a function of the nanobrick side length $L_x$ and air gap $t_a$ with other parameters being as follows: $\lambda = 800$ nm, $t_m = 50$ nm, $\Lambda = 250$ nm and $L_y = L_x$. Coloration is related to the reflection amplitude, while the magenta lines represent constant reflection phase contours. (**C**) Reflection phase (dashed) and amplitude (solid) dependencies on the nanobrick length $L_x$ for two extreme air gaps: $t_a = 20$ (red) and 350 (blue) nm. Circles represent the nanobrick sizes selected for the OMS supercell designed for dynamic beam steering. (**D**) Top view and (**E**) cross section of the designed MEMS-OMS supercell. (**F**, **G**) Distributions of the reflected TM electric field (*x*-component) at 800 nm wavelength for air gaps of $t_a = 20$ and 350 nm, respectively. (**H**) Diffraction efficiencies of different orders ($|m| \leq 1$) calculated as a function of the air gap $t_a$ for TM/TE incident light with 800 nm wavelength. (**I**) Diffraction efficiencies of different orders ($|m| \leq 1$) calculated at the air gap $t_a = 20$ nm as a function of the wavelength for TM/TE incident light.

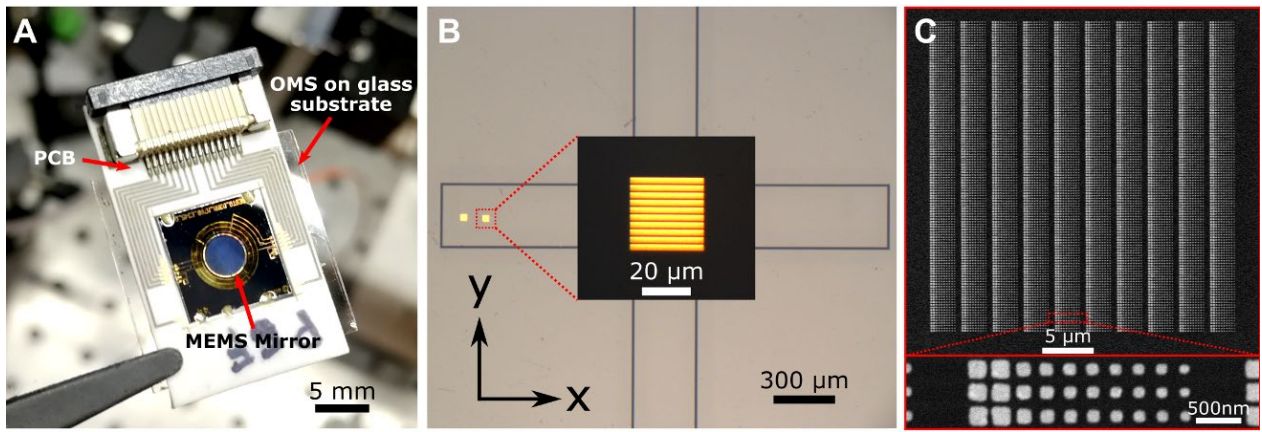

**Fig. 3. MEMS-OMS assembly.** (**A**) Typical photo of the MEMS-OMS assembly consisting of the OMS patterned on a glass substrate, an ultra-flat thin-film MEMS mirror and a printed circuit board (PCB) for electrical connection. (**B**) Optical microscopy and (**C**) SEM images of the OMS representing the 30×30 $\mu m^2$ and 250-nm-period array of differently sized gold nanobricks designed for dynamic beam steering, fabricated atop a 10-$\mu$m-high pedestal on the glass substrate and used in the MEMS-OMS assembly.

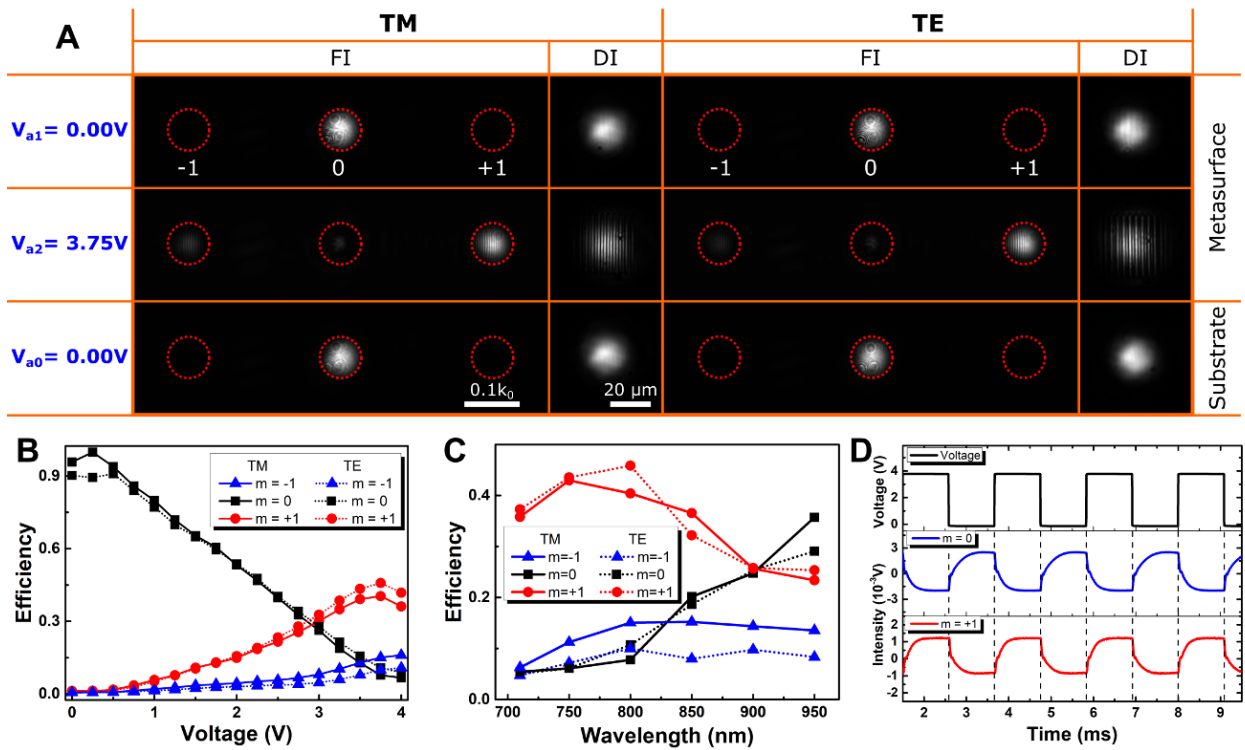

**Fig. 4. Polarization-independent dynamic beam steering: characterization.** (**A**) Optical images at the direct object (DI) and Fourier image (FI) planes of the reflected light from MEMS-OMS under actuation voltages of $V_{a1}$ = 0.00 V (upper panel) and $V_{a2}$ = 3.75 V (middle panel) for TM/TE normally incident light with 800 nm wavelength. Reflected light from unstructured substrate (bottom panel) in the MEMS-OMS device is also recorded as a reference. (**B**) Diffraction efficiencies of different orders ($|m| \leq 1$) measured as a function of the actuation voltage for TM/TE incident light with 800 nm wavelength. (**C**) Diffraction efficiencies of different orders ($|m| \leq 1$) measured as a function of the wavelength for TM/TE incident light. (**D**) Response time of the different diffraction orders (m = 0/+1) measured by actuating the MEMS mirror with a periodic rectangle signal.

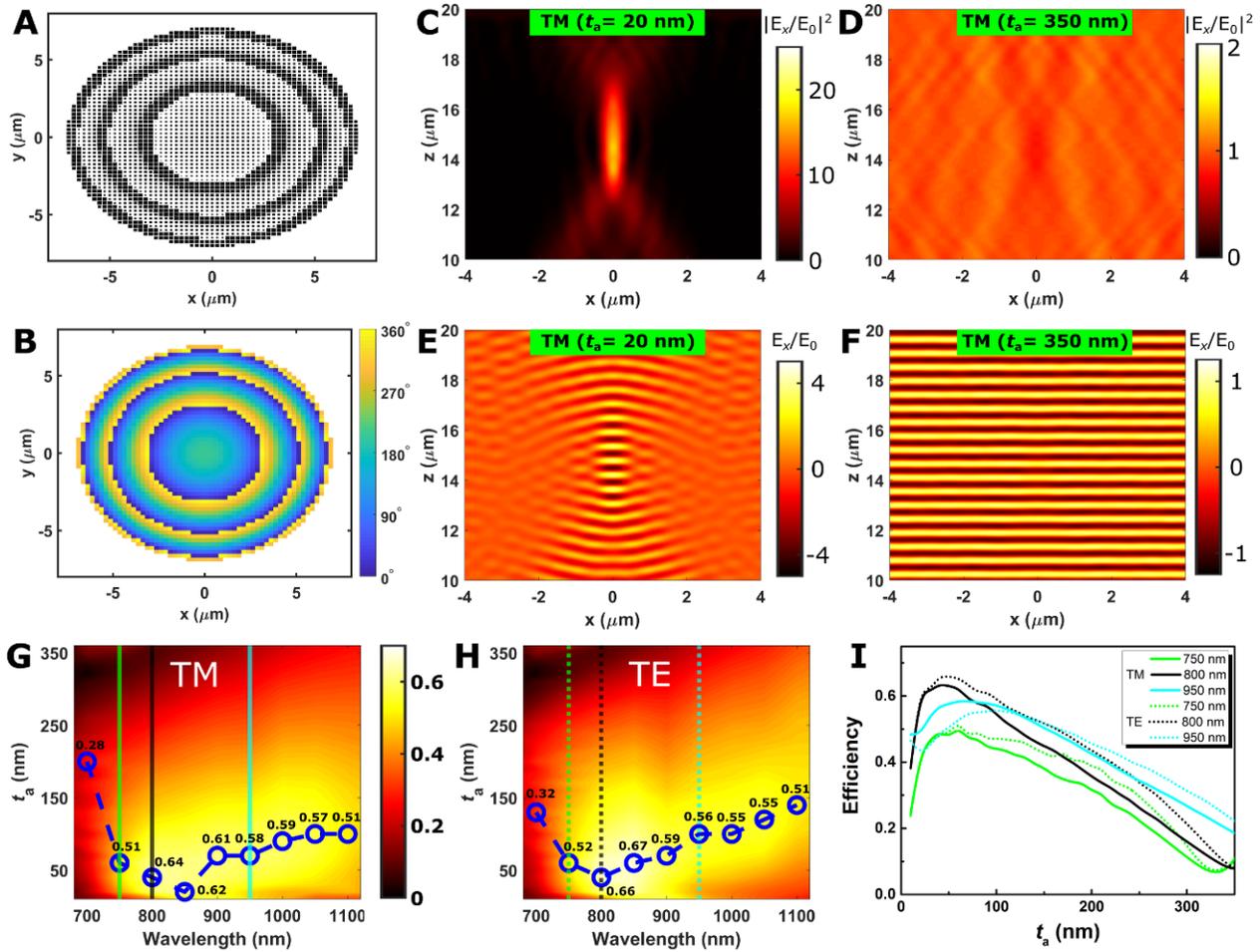

Fig. 5. Polarization-independent dynamic 2D focusing: design. (A) Top view of the OMS designed for dynamic 2D focusing. (B) The phase profile required to focus radiation with focal length of 15 μm at 800 nm wavelength. (C, D) Distributions of the reflected intensity for TM incident light with 800 nm wavelength at air gaps of $t_a$ = 20 and 350 nm, respectively. (E, F) Distributions of the reflected TM electric field (x-component) at 800 nm wavelength for air gaps of $t_a$ = 20 and 350 nm, respectively. (G, H) Focusing efficiencies calculated as a function of the operating wavelength λ and air gap $t_a$, for TM/TE polarizations. The green, black and cyan lines indicate the cases of λ = 750, 800 and 950 nm, respectively. (I) Focusing efficiencies calculated as a function of the air gap $t_a$ for TM/TE polarizations with respective 750, 800 and 950 nm wavelengths.

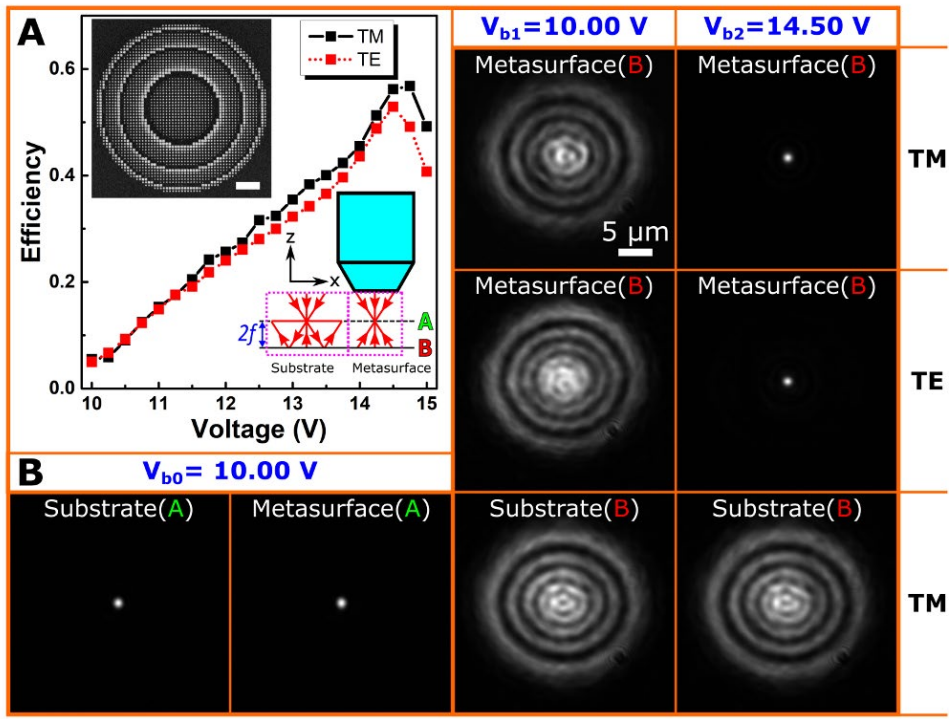

**Fig. 6. Polarization-independent dynamic 2D focusing: characterization.** (**A**) Focusing efficiencies measured as a function of the actuation voltage for TM/TE incident light with 800 nm wavelength. The upper-left inset is a typical SEM image of the OMS representing 14-$\mu$m-diameter and 250-nm-period array of differently sized gold nanobricks designed for dynamic 2D focusing, scalebar 2 $\mu$m. The bottom-right inset illustrate the measurement method in which the incident beam is focused at plane A (focal plane of the objective) and impinging on the unstructured substrate or OMS area of the MEMS-OMS at plane B (2$f$ distance away from the focal plane of the objective), resulting in respective divergent or focused reflected fields. (**B**) Optical images of the reflected light from the unstructured substrate and OMS area of the MEMS-OMS positioned at plane B with actuation voltages of $V_{b1}$ = 10.00 V and $V_{b2}$ = 14.50 V for TM/TE incident light at 800 nm wavelength. The reflected light from the unstructured substrate and OMS area of the MEMS-OMS positioned at plane A was also recorded as a reference.

SUPPLEMENTARY MATERIALS

# Dynamic MEMS-based optical metasurfaces

Chao Meng,[1†] Paul C. V. Thrane,[2†] Fei Ding,[1] Jo Gjessing,[2] Martin Thomaschewski,[1] Cuo Wu[1,3], Christopher Dirdal,[2*] Sergey I. Bozhevolnyi[1*]

[1]Centre for Nano Optics, University of Southern Denmark, Campusvej 55, Odense DK-5230, Denmark.
[2]SINTEF Microsystems and Nanotechnology, Gaustadalleen 23C, 0737 Oslo, Norway.
[3]Institute of Fundamental and Frontier Sciences, University of Electronic Science and Technology of China, Chengdu 610054, China.

[†] These authors contributed equally to this work.
[*] Corresponding author emails: christopher.dirdal@sintef.no (C.D.); seib@mci.sdu.dk (S.I.B.)

**Content**
Section S1. Optimization of nanobrick thickness ($t_m$)
Section S2. Design of the OMS supercell for polarization-independent dynamic beam steering
Section S3. MEMS-OMS assembly and pre-characterizations
Section S4. Experimental setup for measuring MEMS-OMS devices
Section S5. Simulation and fabrication of the MEMS-OMS for polarization-independent dynamic 2D reflective focusing
Section S6. MEMS-OMS for polarization-independent dynamic beam steering among three diffraction orders
Section S7. Dynamic MEMS-OMS based on modulated coupled SR-SPPs/FP resonances
Fig. S1. Optimization of nanobrick thickness ($t_m$)
Fig. S2. Design of the OMS supercell for polarization-independent dynamic beam steering
Fig. S3. MEMS mirror fabrication and MEMS-OMS assembly
Fig. S4. Smallest achievable separation between the MEMS mirror and OMS surface
Fig. S5. Experimental setup for measuring MEMS-OMS devices
Fig. S6. Design and calculated performances of the MEMS-OMS for polarization-independent dynamic beam focusing
Fig. S7. Fabrication of the OMS for polarization-independent dynamic 2D beam focusing
Fig. S8. Polarization-independent dynamic beam steering among three diffraction orders: design
Fig. S9. Polarization-independent dynamic beam steering among three diffraction orders: characterization
Fig. S10. Dynamic MEMS-OMS based on modulated coupled SR-SPPs/FP resonances

## Section S1. Optimization of nanobrick thickness ($t_m$)

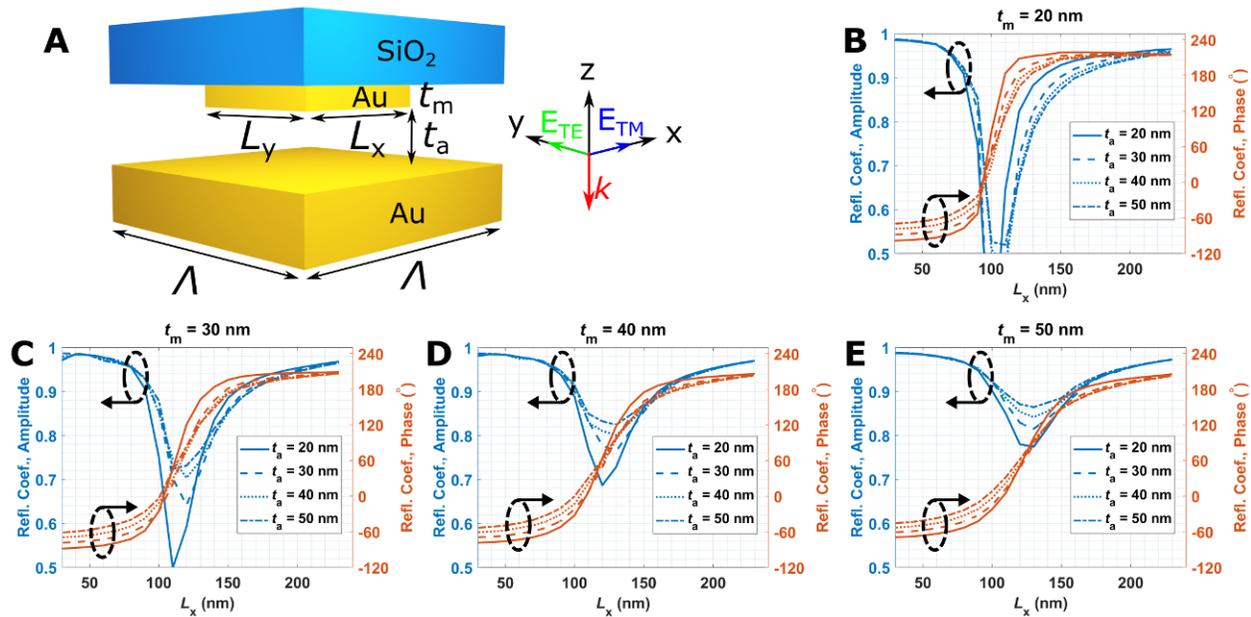

**Fig. S1. Optimization of nanobrick thickness ($t_m$).** (**A**) Schematic of the OMS unit cell including a gold nanobrick on a glass substrate, an air gap and a gold mirror. (**B-E**) Reflection phase (red) and amplitude (blue) dependencies on the nanobrick length $L_x$ for different nanobrick thicknesses $t_m$ (20–50 nm) and air gaps $t_a$ (20–50 nm) under TM incident light, with other parameters being as follows: $\lambda$ = 800 nm, $\Lambda$ = 250 nm and $L_y = L_x$.

## Section S2. Design of the OMS supercell for polarization-independent dynamic beam steering

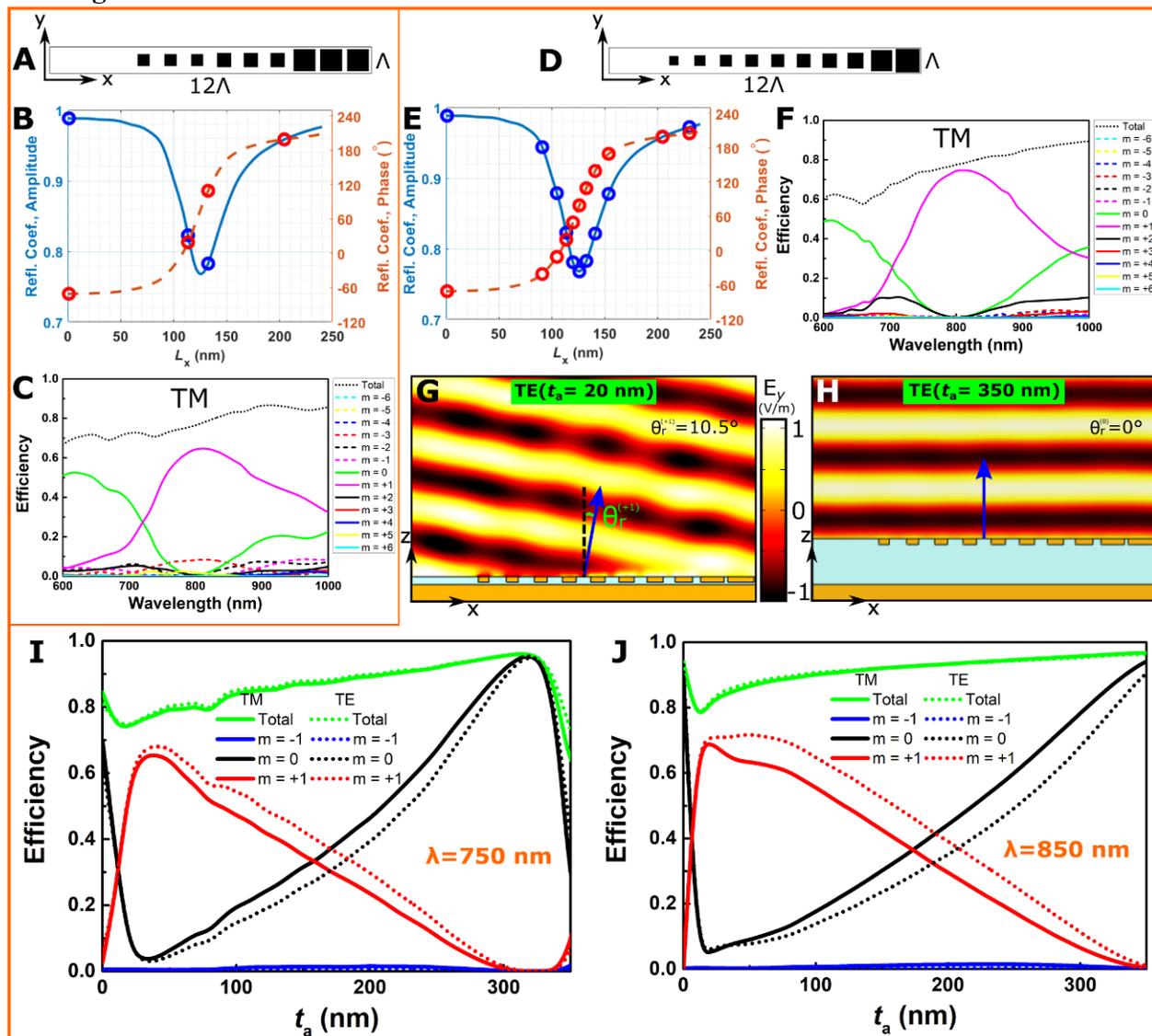

**Fig. S2. Design of the OMS supercell for polarization-independent dynamic beam steering.** (**A-C**) Beam steering OMS designed with 90° phase steps, in which the supercell is composed from triplicated ($\Lambda_{sc}$ = 4×3 = 12$\Lambda$) cells. (**D-J**) Beam steering OMS designed with best approximated phase profile, in which the supercell is composed from extracting unit cells to approximate the required phase profile with smallest deviations. (**A, D**) Top view of the designed OMS supercells. (**B, E**) Reflection phase (red dashed) and amplitude (blue solid) dependencies on the nanobrick length $L_x$ at $t_a$ = 20 nm, with other parameters being as follows: $\lambda$ = 800 nm, $\Lambda$ = 250 nm, $t_m$ = 50 nm, and $L_y = L_x$. Circles represent the nanobrick sizes selected for the OMS supercell designed for dynamic beam steering. (**C, F**) Diffraction efficiencies of different orders calculated as a function of wavelength for TM incident light at an air gap of $t_a$ = 20 nm. (**G, H**) Distributions of the reflected TE electric field (*y*-component) at 800 nm wavelength for air gaps of $t_a$ = 20 and 350 nm, respectively. (**I, J**) Diffraction efficiencies of different orders calculated as a function of the air gap $t_a$ for TM/TE incident light with 750 and 850 nm wavelengths, respectively.

# Section S3. MEMS-OMS assembly and pre-characterizations

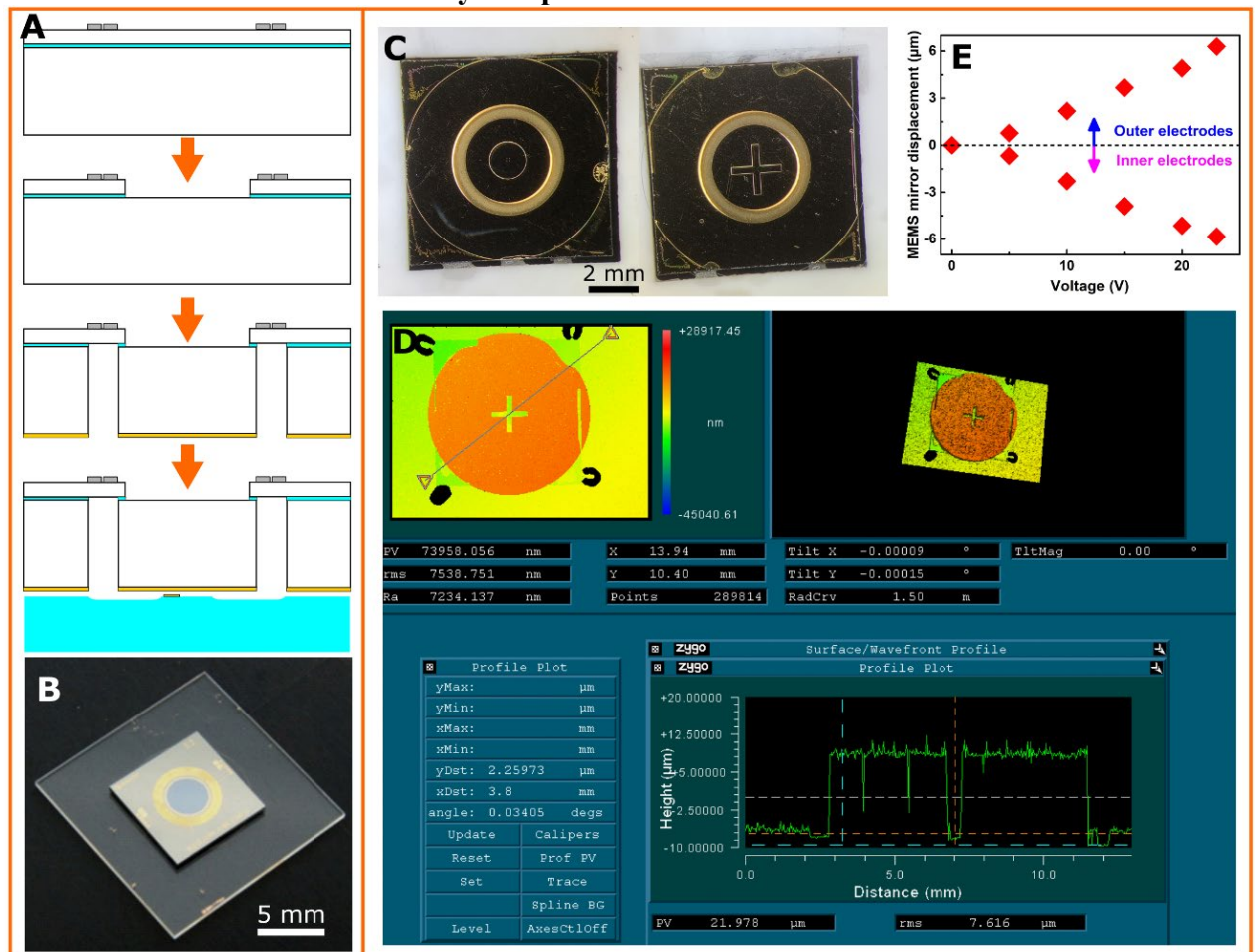

**Fig. S3. MEMS mirror fabrication and MEMS-OMS assembly.** (**A**) MEMS mirror fabrication and MEMS-OMS assembly processes. (**B, C**) Typical photos of the MEMS-OMS assembly before wire-bonding, taken from the respective MEMS-mirror and OMS sides. (**D**) Typical mirror-glass separation (~ 2 μm) measured with unactuated MEMS mirror using a white light interferometry. (**E**) Measured MEMS mirror displacement by applying actuation voltages from 0 to 23 V.

The air gap between the MEMS mirror and glass substrate can be measured by illuminating the MEMS-OMS device with monochromatic light and observing the interference fringes related to the reflection light from the MEMS mirror and the glass substrate. However, from this measurement alone it is not possible to determine which of the fringes correspond to which mirror-glass separation. Therefore, light with a second wavelength is used to make another group of fringes that are shifted and have a different periodicity with respect to the first fringe pattern, as shown in fig. S4, A and B. Here we used a 532 nm laser and a 492 nm filtered thermal light, which interfere destructively at a mirror-glass separation of around 1.7 μm (magenta mark in fig S4). Thus, by finding the MEMS mirror position that gives destructive interference it is possible to label each fringe with a single separation distance, as long as the original separation is known to within approximately 4 μm. In practice, this process was done by first tilting the MEMS mirror such that many fringes were visible at once (fig. S4, A-C), before bringing the mirror parallel while keeping track of which fringes were visible (fig. S4D). For several assemblies, the smallest achievable mirror-glass separation ($t_a+t_m$) by actuated MEMS mirror could be as small as ~ 100 nm (yellow marks in fig. S4), corresponding to $t_a$ ~ 50 nm, and these samples were then selected for further optical characterizations.

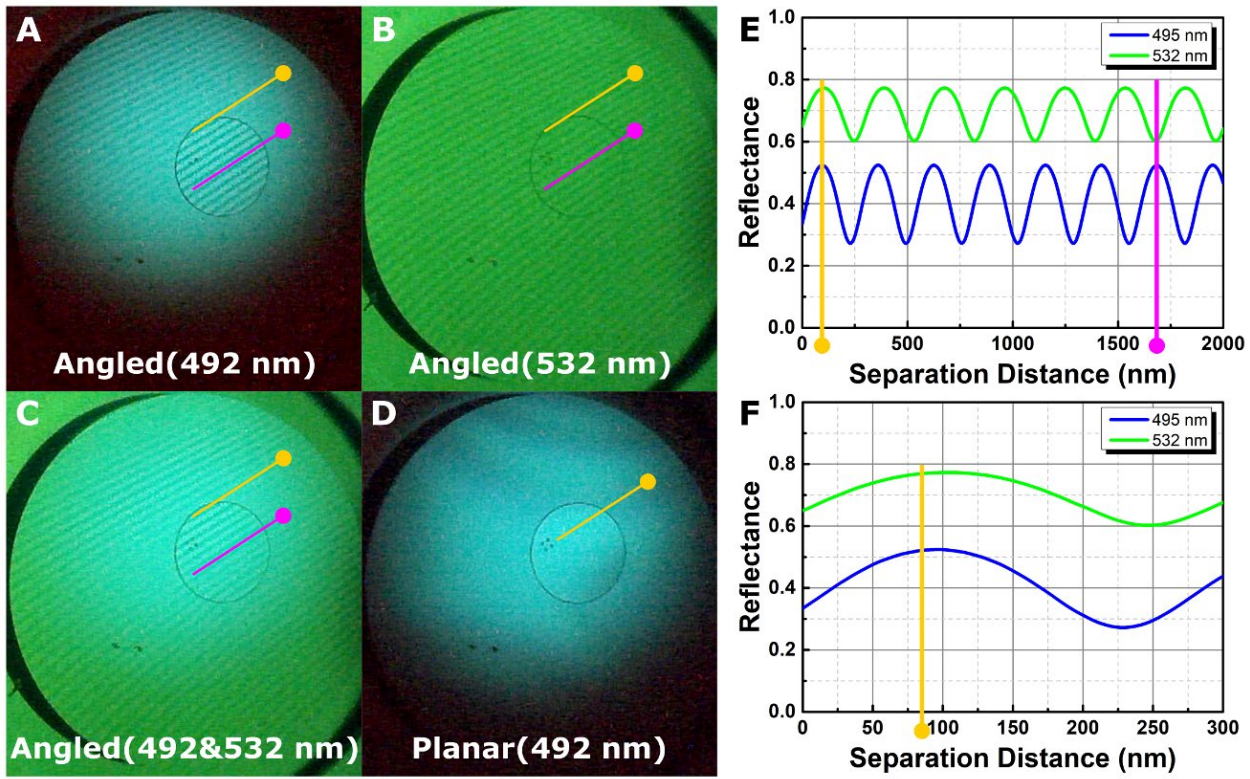

Fig. S4. Smallest achievable separation between the MEMS mirror and OMS surface. (A-C) Typical optical images of the interference patterns with slightly tilted MEMS mirror under (A) 492-nm, (B) 532-nm, (C) 492 and 532-nm incident light. (D) Typical optical images of the interference patterns with bringing MEMS mirror parallel to the glass substrate under 492-nm incident light. (E) The interference fringes calculated as a function of the spacing between the MEMS mirror and the glass substrate ($t_a + t_m$), while the magenta and yellow makers indicate the identified interference fringes corresponding to mirror-glass spacing of ~ 1.7 $\mu$m and ~ 100 nm, respectively. (F) Zoomed-in view of (E), indicating the smallest achievable mirror-glass ($t_a + t_m$) separation of ~ 100 nm, corresponding to $t_a$ ~ 50 nm.

## Section S4. Experimental setup for measuring MEMS-OMS devices

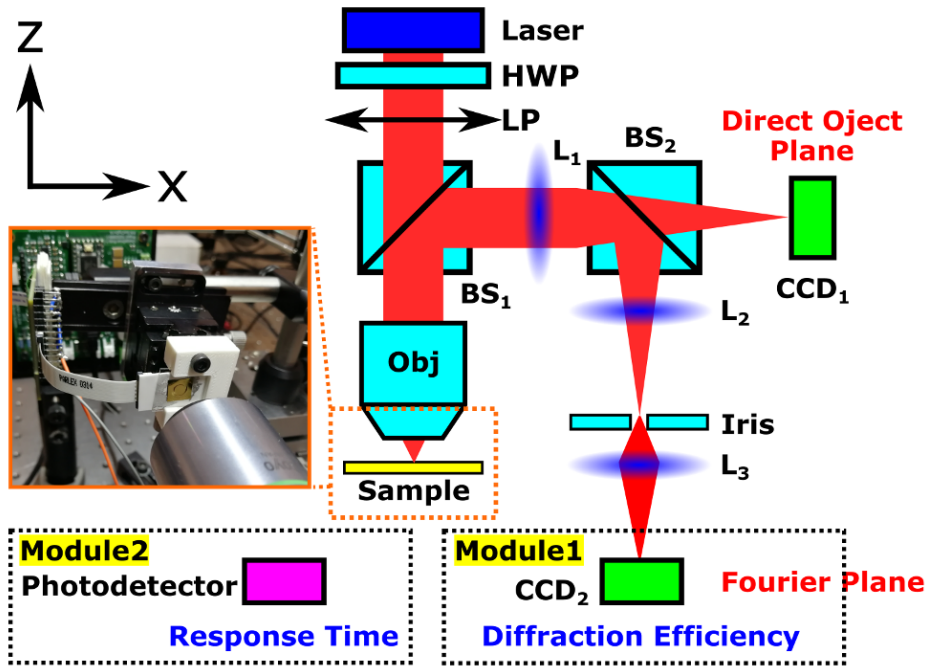

**Fig. S5. Experimental setup for measuring MEMS-OMS devices.** Laser: Ti:sapphire laser (Spectra Physics 3900S); HWP: Half-wave plate; LP: Glan-Thomson polarizer; $BS_{1/2}$: Beam splitter; Obj: Objective (Mitutoyo, M Plan Apo 20×/0.42 and 50×/0.55 for measuring respective MEMS-OMSs for dynamic beam steering and focusing); $L_1$: tube lens ($f$ = 200 mm); $L_{2/3}$: convex lens ($f$ = 100 mm). Modules 1 and 2 are used to estimate the diffraction efficiencies and response time of the MEMS-OMS for dynamic beam steering, respectively.

**Section S5. Simulation and fabrication of the MEMS-OMS for polarization-independent dynamic 2D reflective focusing**

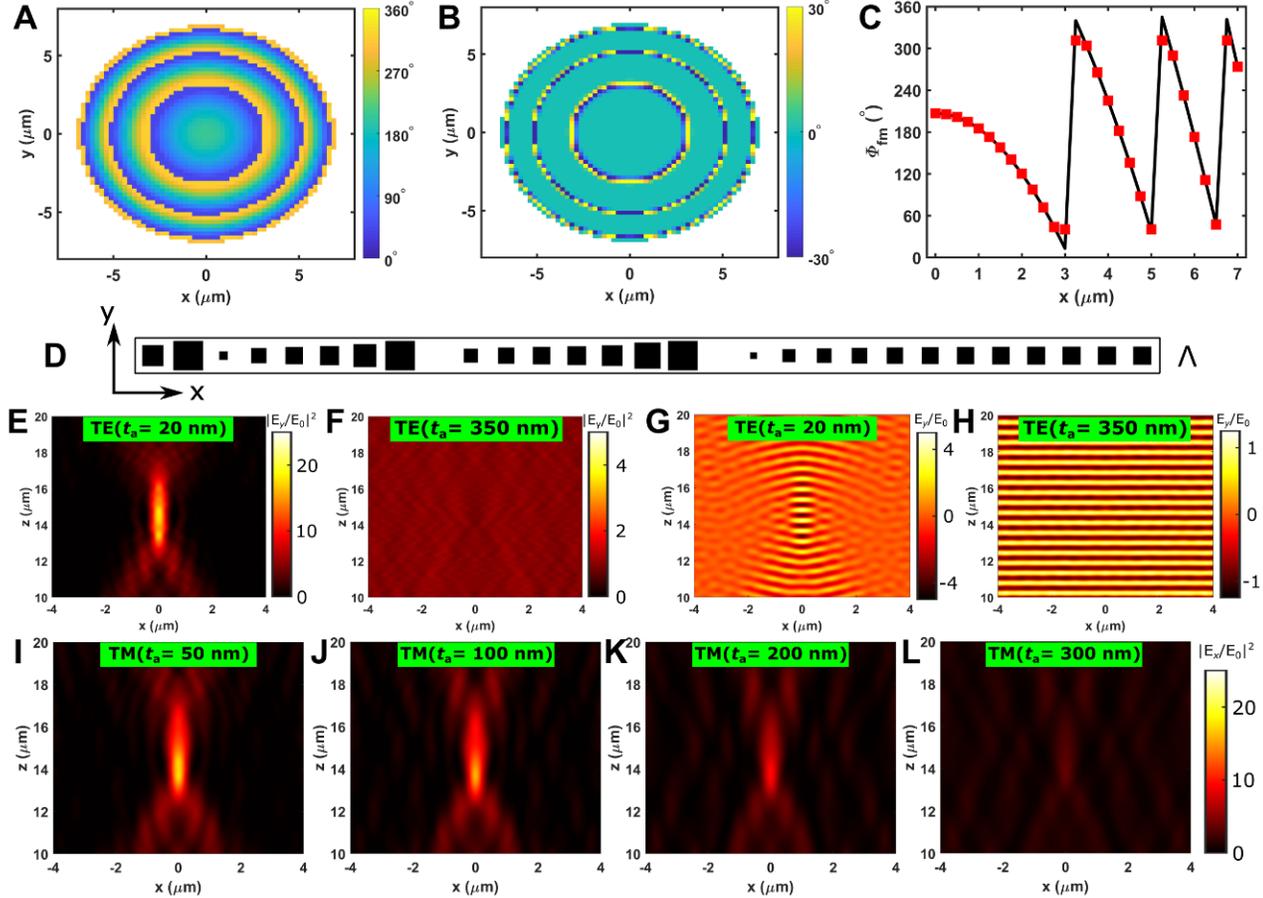

**Fig. S6. Design and calculated performances of the MEMS-OMS for polarization-independent dynamic beam focusing.** (**A**) Phase profile of the designed OMS constructed by the available nanobricks with an air gap of $t_a$ = 20 nm at 800 nm wavelength. (**B**) Deviations between the required (Fig. 5A) and available phase profiles. (**C**) The phase profile required for 1D beam focusing OMS with a focal length $f$ = 15 $\mu$m at 800 nm wavelength, while the red squares indicate the available reflection phases by selecting differently sized nanobricks with an air gap of $t_a$ = 20 nm. (**D**) Top view of half the designed OMS for dynamic 1D beam focusing. (**E, F**) Distributions of the reflected intensity for TE incident light with 800 nm wavelength at air gaps of $t_a$ = 20 and 350 nm, respectively. (**G, H**) Distributions of the reflected TE electric field (y-component) with 800 nm wavelength at air gaps of $t_a$ = 20 and 350 nm, respectively. (**I-L**) Distributions of the reflected intensity for TM incident light with 800 nm wavelength at air gaps of $t_a$ = 50, 100, 200 and 300 nm, respectively.

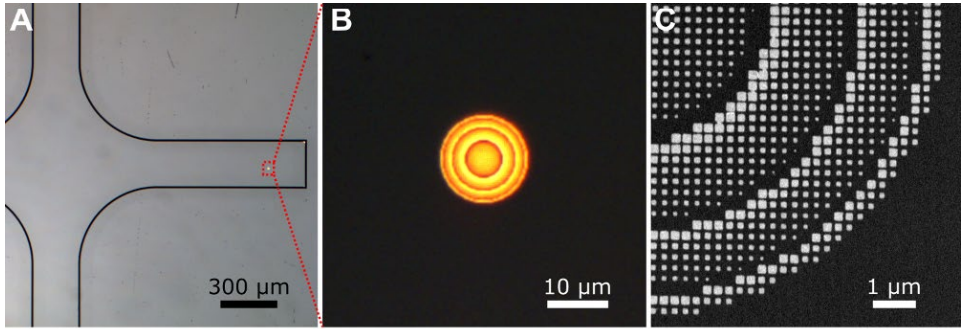

**Fig. S7. Fabrication of the OMS for polarization-independent dynamic 2D beam focusing.** (**A, B**) Optical microscopy and (**C**) SEM images of the OMS representing the 14-μm-diameter and 250-nm-period array of differently sized gold nanobricks designed for dynamic beam focusing, fabricated atop a 10-μm-high pedestal on the glass substrate and used in the MEMS-OMS assembly.

**Section S6. MEMS-OMS for polarization-independent dynamic beam steering among three diffraction orders**

In the main text, the MEMS-OMS for polarization-independent dynamic beam steering between discrete $0^{th}/+1^{st}$ orders (corresponding to steering angle of 0º and 15.5º, in air) have been demonstrated. Moreover, motivated by the nontrivial modification of the size-dependent phase response with varied air gaps and the fine-tuning capabilities (with ~ 1 nm resolution) of the actuated MEMS mirror, we also designed and experimentally demonstrated the MEMS-OMS for polarization-independent dynamic beam steering toward more directions, that is, switching the reflected light among 3 discrete ($0^{th}$, $+1^{st}$, and $+2^{nd}$) diffraction orders (corresponding to steering angles of 0º, 7.7º and 15.5º in air) under normally incident light, enabled by finely actuating the MEMS mirror.

To design the suggested functionality, we resorted to a shared-aperture strategy (*59*), in which different OMS designed with distinct functionalities are randomly interleaved, enabling thus more versatile functionalities and/or multiplexing channels within single OMS. Different from the conventional randomly interleaved OMS, our OMS configuration was designed based on two OMSs with different supercells with $\Lambda_{sc1} = 12\Lambda$ and $\Lambda_{sc2} = 24\Lambda$ optimized at two distinct air gaps and then randomly mixed (*59*), promising the possibility of redirecting the reflected light to different ($0^{th}$, $+1^{st}$, and $+2^{nd}$) diffraction orders at different air gaps, with actuated MEMS mirror. One of the basic beam steering OMS ($\Lambda_{sc1} = 12\Lambda$) is exactly the same as that in Fig. 2, which is optimized at $t_a = 20$ nm, while a second beam steering OMS ($\Lambda_{sc2} = 24\Lambda$) is optimized at $t_a = 100$ nm. The second beam steering OMS was then designed by approximating the complex reflection coefficient of an ideal blazed grating with grating (supercell) period of $\Lambda_{sc2}$ (fig. S8, A-C). The reflected electric field (*x*-component) calculated for thus-designed second beam steering OMS under TM incident light at 800 nm wavelength with $t_a = 100$ and 350 nm manifest smooth wavefronts travelling in the direction of the $+1^{st}$ and $0^{th}$ diffraction orders (corresponding to steering angles of 7.7º and 0º in air) (fig. S8, D and E), promising large modulation efficiencies available with the actuated MEMS-OMS (fig. S8, F and G).

After the design of the second beam-steering OMS, we construct the overall MEMS-OMS for dynamic beam steering among three diffraction orders by randomly interleaving the two basic beam-steering OMSs (optimized at distinct $t_a$ of 20 and 100 nm), as shown in fig. S8H. The dynamic diffraction efficiencies were estimated by simply adding the corresponding diffraction efficiencies of the two basic beam-steering OMSs at different air gaps $t_a$ (Fig. 2H and fig. S8F), according to desired steering directions, showing that the $+2^{nd}/+1^{st}/0^{th}$ diffraction orders (corresponding to steering angles of 15.5º, 7.7º and 0º in air) are dominating at distinct air gaps of $t_a$ around 30, 150 and 350 nm, respectively (fig. S8I).

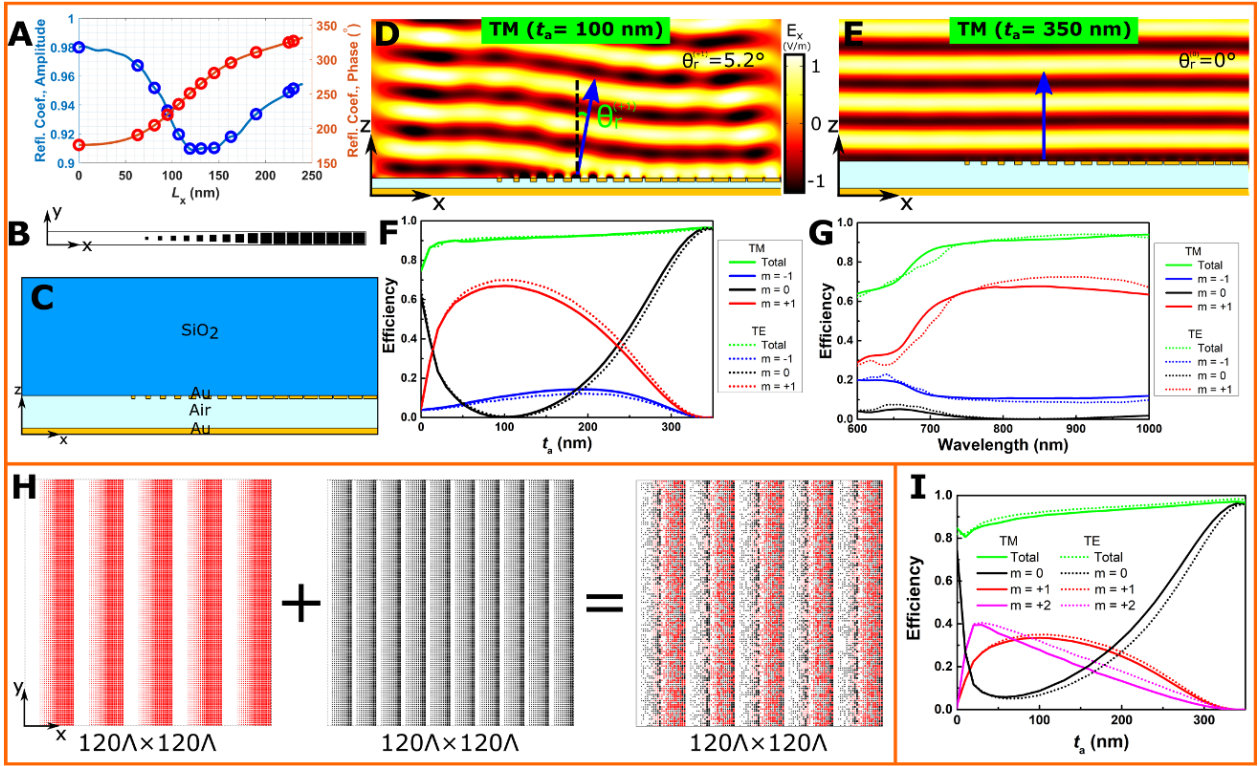

**Fig. S8. Polarization-independent dynamic beam steering among three diffraction orders: design.** (**A**) Reflection phase (red) and amplitude (blue) dependencies on the nanobrick length $L_x$ for $t_a = 200$ nm with other parameters being as follows: $\lambda = 800$ nm, $t_m = 50$ nm, $\Lambda = 250$ nm and $L_y = L_x$. Circles represent the nanobrick sizes selected for the second basic OMS supercell. (**B**) Top view and (**C**) cross section of the second basic OMS supercell with $\Lambda_{sc2} = 24\Lambda$. (**D, E**) Distributions of the reflected TM electric field (x-component) at 800 nm wavelength for air gaps of $t_a = 100$ and 350 nm, respectively. (**F**) Diffraction efficiencies of different orders ($|m| \leq 1$) of the second basic OMS supercell calculated as a function of the air gap $t_a$ for TM/TE incident light with 800 nm wavelength. (**G**) Diffraction efficiencies of different orders ($|m| \leq 1$) of the second basic OMS supercell calculated at an air gap of $t_a = 100$ nm as a function of the wavelength for TM/TE incident light. (**H**) OMS constructed from randomly interleaving two basic beam steering OMS with different supercells with $\Lambda_{sc1} = 12\Lambda$ and $\Lambda_{sc2} = 24\Lambda$ optimized at two distinct air gaps of $t_a = 20$ and 100 nm, respectively. (**I**) Diffraction efficiencies ($m \leq 2$) of MEMS-OMS estimated as a function of the air gap $t_a$ for TM/TE incident light with 800 nm wavelength.

With the above numerical simulations illustrating the possibility of realizing dynamic beam steering among 3 diffraction orders with the designed MEMS-OMS, we now move to the experimental verifications. Following the same fabrication processes detailed in the main text, we fabricated the OMS atop a pedestal on the glass substrate (fig. S9, A and B) and then made the MEMS-OMS assembly. To characterize the MEMS-OMS performance, we electrically actuated the MEMS mirror and monitored the modulated optical response of the MEMS-OMS in both direct object (OMS surface) and Fourier image planes (fig. S9C). In the direct object images, this effect of power redistribution could be identified from the well-pronounced interference fringes formed due to the interference between the residual specular reflection and the +1st ($V_{c2} = 17.00$ V) or +2nd ($V_{c3} = 17.75$ V) order diffracted beams, thus having different periodicities. For both polarizations, the redistribution of the radiated power among the 0th, +1st and +2nd diffraction orders could be visualized (fig. S9, C and D): with the actuation voltage increasing, the first and second diffraction orders became visible, succeeding one another, in accordance with our simulations (fig. S8I).

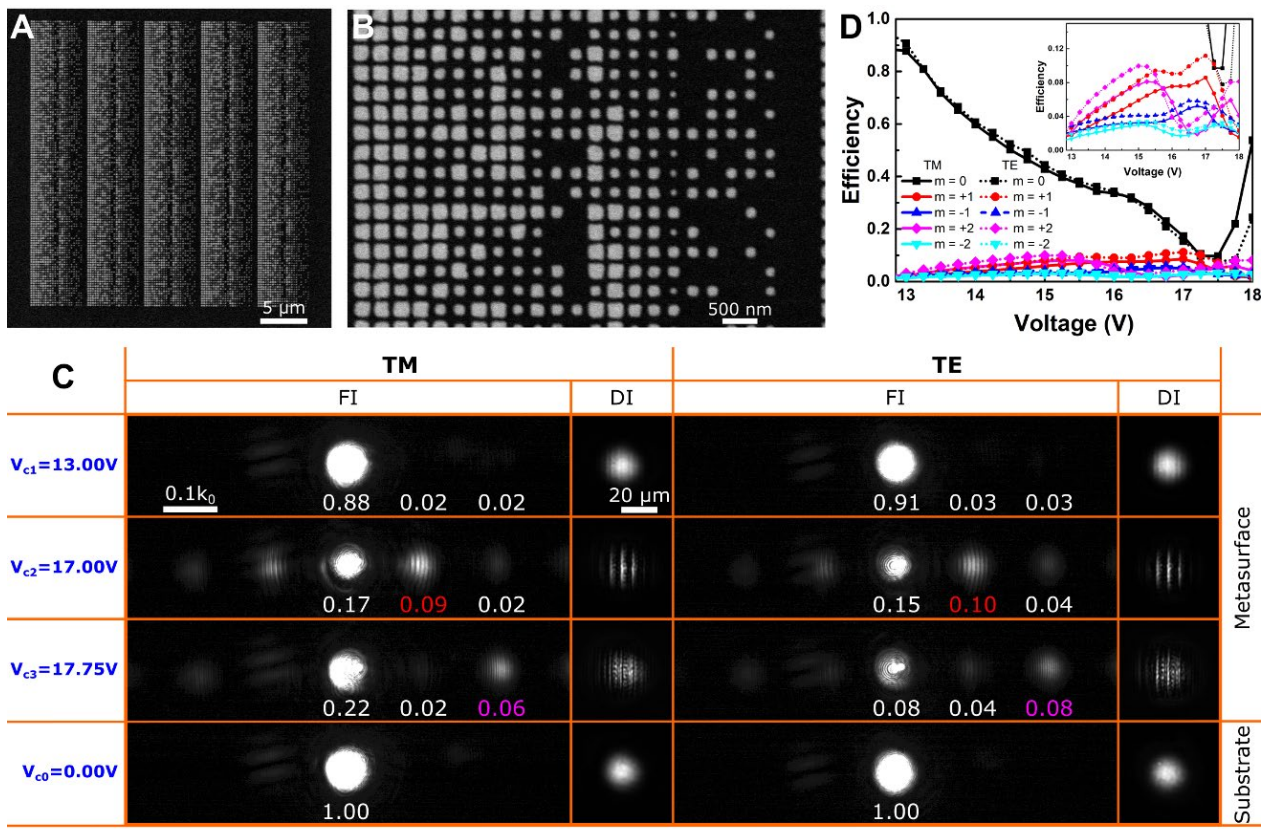

**Fig. S9. Polarization-independent dynamic beam steering among three diffraction orders: characterization.** (**A, B**) SEM images of the OMS representing the 30×30 $\mu m^2$ and 250-nm-period array of randomly interleaved gold nanobricks designed for dynamic beam steering among three diffraction orders, fabricated atop a 10-$\mu$m-high pedestal on the glass substrate. (**C**) Optical images at the direct object (DI) and Fourier image (FI) planes of the reflected light from MEMS-OMS under actuation voltages of $V_{c1}$ = 13.00 V (upper panel), $V_{c2}$ = 17.00 V (upper middle panel) and $V_{c2}$ = 17.75 V (lower middle panel) for TM/TE normally incident light with 800 nm wavelength. Reflected light from unstructured substrate (bottom panel) in the MEMS-OMS device is also recorded as a reference. (**D**) Measured diffraction efficiencies ($|m| \leq 2$) as a function of the actuation voltage for TM/TE incident light with 800 nm wavelength.

**Section S7. Dynamic MEMS-OMS based on modulated coupled SR-SPPs/FP resonances**

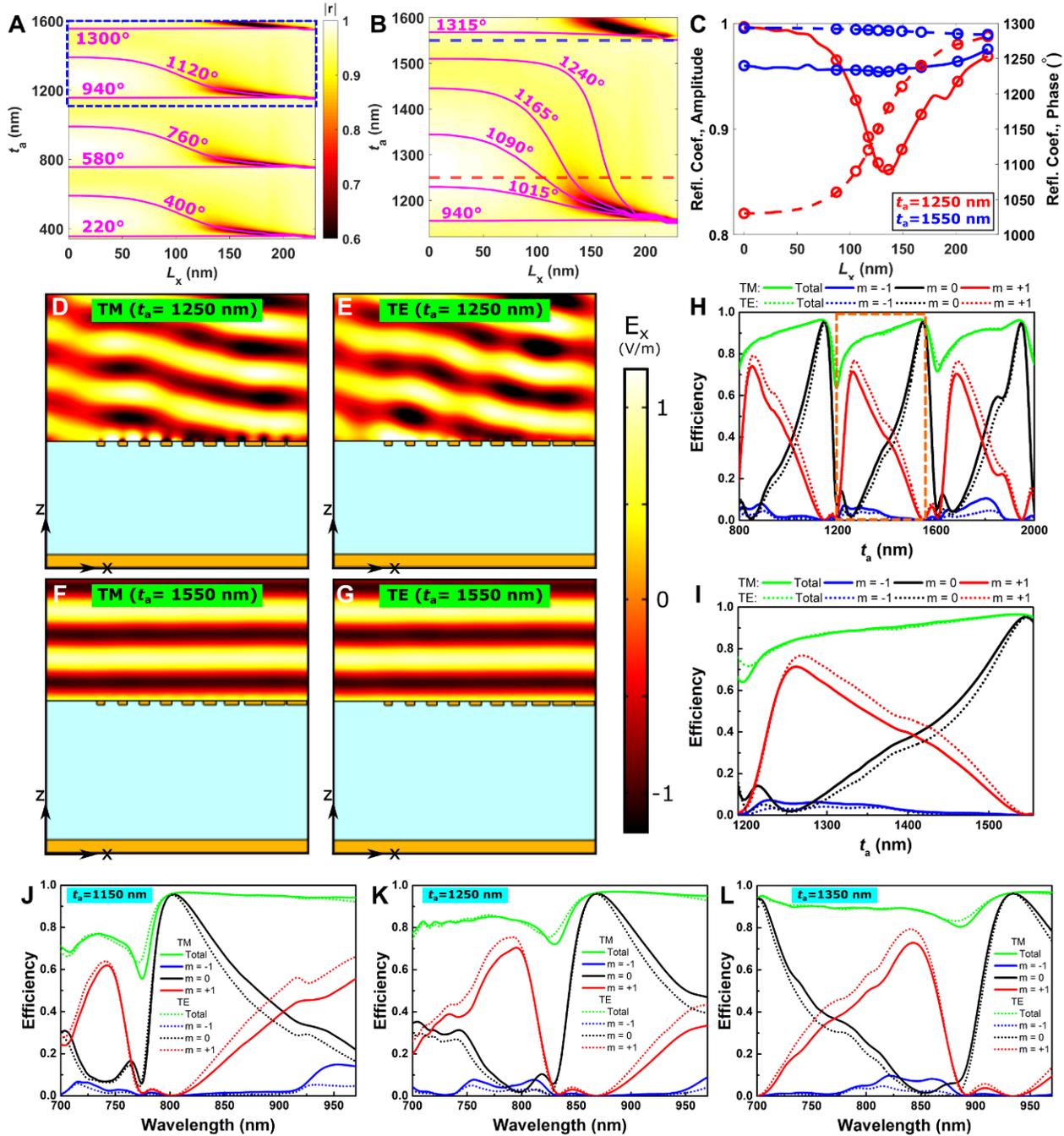

**Fig. S10. Dynamic MEMS-OMS based on modulated coupled SR-SPPs/FP resonances. (A, B)** The complex reflection coefficient $r$ calculated as a function of the nanobrick side length $L_x$ and air gap $t_a$, with $t_a$ varying from **(A)** 340 to 1600 nm and **(B)** 1120 to 1600 nm, while other geometric parameters are as follows: $\lambda$ = 800 nm, $t_m$ = 50 nm, $\Lambda$ = 250 nm and $L_y = L_x$. Coloration is related to the reflection amplitude, while the magenta lines represent constant reflection phase contours. **(C)** Reflection phase (dashed) and amplitude (solid) dependencies on the nanobrick length $L_x$ for two extreme air gaps: $t_a$ = 1250 (red) and 1550 (blue) nm. Circles represent the nanobrick sizes selected for the OMS supercell designed for dynamic beam steering. **(D-G)** Distributions of the reflected TM/TE electric field ($x/y$-component) at 800 nm wavelength for air gaps of $t_a$ = **(D, E)** 20 and **(F, G)** 350 nm, respectively. **(H, I)** Diffraction efficiencies of different orders ($|m| \leq 1$) calculated as a function of the air gap $t_a$ for TM/TE incident light with 800 nm wavelength, with varying $t_a$ from **(H)** 800 to 2000 nm and **(I)** 1200 to 1580 nm, respectively.

**(J-K)** Diffraction efficiencies of different orders ($|m| \leq 1$) calculated as a function of the wavelength for TM/TE incident light, with air gaps of $t_a =$ **(J)** 1150, **(K)** 1250 and **(L)** 1350 nm.